\documentclass[aps,showkeys,superscriptaddress,byrevtex,a4paper,12pt]{revtex4}
\usepackage{dcolumn}
\usepackage{bm}
\usepackage[latin2]{inputenc}
\usepackage{graphics}
\usepackage{longtable}
\usepackage{amsmath}
\usepackage{amssymb}
\usepackage{enumerate}

\begin{document}


\title{ Reduced density matrices, their spectral resolutions, and the Kimball-Overhauser approach}

\author{P. Ziesche}
\email[E-mail: ]{pz@mpipks-dresden.mpg.de}
\affiliation{Max-Planck-Institut f\"ur Physik komplexer Systeme,
  N\"othnitzer Str. 38, D-01187 Dresden, Germany}
\date{\today}

\author{F. Tasn\'adi}
\email[E-mail: ]{f.tasnadi@ifw-dresden.de}
\affiliation{Leibniz-Institut f\"ur Festk\"orper- und Werkstoffforschung, \\
Helmholtzstr. 20, D-01069 Dresden, Germany, and \\
University of Debrecen, Hungary}


\begin{abstract}
Recently, it has been shown, that the pair density of the homogeneous
electron gas can be parametrized in terms of 2-body wave functions (geminals),
which are scattering solutions of an effective 2-body Schr\"odinger equation.
For the corresponding scattering phase shifts, new sum rules are reported 
in this paper. These sum rules describe not only the normalization of the pair 
density (similar to the Friedel sum rule of solid state theory), but also the 
contraction of the 2-body reduced density matrix. This allows one to calculate 
also the momentum distribution, provided that the geminals are known from an 
appropriate screening of the Coulomb repulsion. An analysis is presented leading
from the definitions and (contraction and spectral) properties of reduced 
density matrices to the Kimball-Overhauser approach and its generalizations.
Thereby cumulants are used. Their size-extensivity is related to the 
thermodynamic limit.  
 
\end{abstract}
%
\maketitle
\section*{Introduction}

The description of an electronic many-body system in its ground-state (GS) by 
means of 1-body orbitals $\psi^{}_{\kappa}(1)$ with the short-hand $1\equiv
(\mathbf{r}_1,\sigma_1)$ is well-known and widely used. For example in the 
configuration-interaction (CI) method \cite{Ful,Bart,Kutz1} these orbitals are
used to build up the $N$-electron Slater determinants. The linearly combined 
Slater determinants form the GS-wave function $\Psi(1,\ldots ,N)$ such that the 
electron density is given by $\rho(1)=\sum_{\kappa}\nu_{\kappa}|
\psi^{}_{\kappa}(1)|^2$ with a non-idempotent occupancy
$\nu_{\kappa}$ being between 0 and 1 as a consequence of the Pauli principle,
$\sum_\kappa\nu_\kappa=N$. In the
Hartree-Fock approximation and in the Kohn-Sham treatment of the 
density-functional theory (DFT), cf. e.g. \cite{Esch}, the orbitals $\psi^{}_{\kappa}(1)$ are solutions of effective 1-body
Schr\"odinger equations and the occupanies are idempotent, $\nu_{\kappa}^2=\nu_{\kappa}$,
using thereby the aufbau principle. The DFT shows the `power and charme' of the 
1-body density $\rho(1)$ \cite{Levy}. In the theory of reduced density matrices 
(RDMs) \cite{Davi,Harri,Erd,Cio,Cole,Vald,Kutz2} and in the density matrix 
functional theory (DMFT) \cite{Cio,Cio1} the
$\psi^{}_{\kappa}(1)$ diagonalize the 1-body reduced density matrix (1-matrix) 
so that $\gamma(1|1')=\sum_{\kappa}\psi^{}_{\kappa}(1)\nu_{\kappa}
\psi^{\ast}_{\kappa}(1')$ holds, what is called spectral resolution. The 
$\nu_{\kappa}$ are again non-idempotent as a consequence of the electron
correlation phenomenon such that the non-idempotency of $\gamma$ or 
$\nu_{\kappa}$ can be used to measure the strength of correlation, 
$\nu_\kappa^2<\nu_\kappa$, ${\rm Tr}\gamma^2<{\rm Tr}\gamma$ 
\cite{Zie1,Zie2,Zie3,Roe}. For crystalline solids the $\nu_\kappa$ describe the 
occupation band structure. The $\psi_\kappa(1)$ are the natural orbitals and 
the $\kappa$ are 1-body quantum numbers.

The 1-matrix $\gamma(1|1')$ is a 2-point function, not to be confused 
with another important 2-point function, the pair density (PD) $\rho_2(1,2)$. 
This PD shows its power and charm in the Fermi hole for parallel-spin pairs (due
to the Pauli `repulsion' and modified by the Coulomb repulsion), in the Coulomb hole for antiparallel-spin pairs (due to the Coulomb
repulsion), in the coalescing (or on-top) cusp and curvature properties, and in
the possibility to calculate and discuss particle number fluctuations
in partial regions of the system (e.g. Daudel loges or "stockholder" loges or 
Bader basins or Wigner-Seitz cells or
$\cdots$) as another (more sensitive) measure of the correlation strength with
the conclusion `(strong) correlations (strongly) suppress such fluctuations', 
cf. \cite{Ful}, p. 157, and \cite{Roe,Zie8}. Strong electron correlation
thus localizes electrons. Examples are the Wigner crystallization of the
low-density HEG, the Wigner-like charge ordering in Yb$_4$As$_3$ \cite{Ful2},
and the insulating GS of CoO (where DFT predicts a metal).

Related to the PD is the less known and less worked out use of 2-body wave 
functions (geminals) $\psi^{}_K(1,2)$ for the description of an electronic 
many-body system \cite{Sur}. The $K$ are 2-body quantum numbers. 
Recently, in an eventually possible 
pair density functional theory (PDFT) \cite{Zie4} and in the Kimball-Overhauser
approach of the homogenous electron gas (HEG) \cite{Kim,Over,Gor,Dav1,Dav2,Dav3,
Zie5} such geminals are discussed as solutions of an effective 
2-body Schr\"odinger equation and used to parametrize the pair density (PD) 
$\rho_2(1,2)=\sum_K\mu_K|\psi^{}_K(1,2)|^2$. More general, the so-called 
natural geminals diagonalize the 2-body RDM (2-matrix) $\gamma_2(1|1',2|2')=
\sum_K\psi^{}_K(1,2)\mu_K\psi^{\ast}_K(1',2')$. It is quite natural to CI 
expand the $\psi_K(1,2)$ in terms of the $\psi_\kappa(1)$. In Ref. \cite{Zie6} 
it is suggested to generalize the Kimball-Overhauser approach by using its PD 
geminals as natural geminals in the spectral resolution of $\gamma_2$. This 
generalization has the advantage that from the natural geminals not only follows
the PD, but also the 1-matrix $\gamma$ using thereby the contraction properties
of $\gamma_2$. - Geminals appear also in "the
antisymmetrized power (AGP) function as a flexible ansatz for fermion systems
with arbitrary $N$" \cite{Cole} and they appear in a generalized DFT in
2-particle space, cf. \cite{Gon}, p. 325 and refs. therein. Besides, there are 
links between DMFT and geminal theory \cite{Cio2}. Finally, a general
remark of Davidson is "scientists have not yet learned to think in terms of
$\binom{N}{2}$ geminals rather than in $N$ orbitals" \cite{Davi}, p. 97. 

In the following, the Kimball-Overhauser approach is summarized from an RDM 
point of view and possible generalizations are presented.
%
\section*{Basic notation}

The GS energy $E$ of a non-relativistic $N$-electron system in Born-Oppenheimer
approximation is partitioned as $E=T[\gamma]+V_{\rm ext}[\rho]+V_{\rm int}
[\rho_2]$, where the well-known linear functionals (using the short-hand 
notation $\int d1\equiv\sum_{\sigma_1}\int d^3r_1$)
\begin{eqnarray}
T[\gamma]=\int d1 \ t(\mathbf{r}_1)\gamma(1|1')|_{1'=1}, \quad
t(\mathbf{r}_1)=-\frac{\hbar^2}{2m}\left(\frac{\partial}
{\partial\mathbf{r}_1}\right)^2, \nonumber \\
V_{\rm ext}[\rho]=\int d1 \ \rho(1)v_{\text{ext}}(\mathbf{r}_1), \quad 
V_{\text{int}}[\rho_2]=
\int \frac{d1d2}{2!}\ \rho_2(1,2)\frac{\epsilon^2}{r_{12}}, \quad \epsilon^2=\frac{e^2}{4\pi \varepsilon_0}
\label{energies}
\end{eqnarray}
appear. $v_{\text{ext}}(\mathbf{r})$ is the potential of the nuclei (or the jellium background) binding
the electrons. The 1-matrix $\gamma(1|1')$, the electron density $\rho(1)\ge 0$ and the PD
$\rho_2(1,2)\ge 0$ are functionals of the GS wave function $\Psi(1,\ldots,N)$ arising from
$\Psi^{}(1,\ldots,N)\Psi^{\ast}(1',\ldots,N')$ by appropriate contractions, 
which means the operations $i'=i$ and $\int di$. (In the many-body perturbation theory this term has
another meaning.) $\gamma$, $\rho$ and $\rho_2$ follow from the 2-matrix (which
is defined
by the $N-2$ contraction of $\Psi^{}(1,\ldots,N)\Psi^{\ast}(1',\ldots,N')$),
\begin{equation}
\gamma_2(1|1',2|2')=\int \frac{d3\ldots dN}{(N-2)!}\Psi^{}(1,2,3,\ldots)
\Psi^{\ast}(1',2',3,\ldots), \quad \text{Tr}\gamma_2=N(N-1), 
\end{equation}
where the wave-function normalization
\begin{equation}
\int \frac{d1\ldots dN}{N!}|\Psi^{}(1,\ldots,N)|^2=1
\end{equation}
is used (each particle configuration is naturally counted only once).
%
\section*{Contraction sum rules, cumulant expansions, spin structures, and pair 
densities}

The contraction of the 2-matrix yields
\begin{equation}
\gamma(1|1')=\frac{1}{N-1}\int d2 \ \gamma_2(1|1',2|2), \quad 
\rho(1)=D\gamma(1|1')=\gamma(1|1).
\label{contrSR}
\end{equation}
D projects out the off-diagonal elements $i'\neq i$. 
The contraction of $\rho_2=D\gamma_2$ yields 
\begin{equation}
\rho(1)=\frac{1}{N-1}\int d2 \ \rho_2(1,2).
\end{equation}
${\rm Tr}\gamma=N$ follows from ${\rm Tr}\gamma_2=N(N-1)$. $\rho$ is needed for
the expectation value of the external (the electrons confining) potential, 
whereas $\gamma$ and $\rho_2$ are the quantities to calculate the kinetic and 
the interaction energy, respectively, cf. Eq.(\ref{energies}).

A fairly natural partitioning of $\gamma_2$ is its cumulant expansion
\begin{eqnarray}
\gamma^{}_2(1|1',2|2')=\gamma^{\text{HF}}_2(1|1',2|2')-\chi(1|1',2|2'), \nonumber \\
\gamma^{\text{HF}}_2(1|1',2|2')=A\gamma(1|1')\gamma(2|2')=\gamma(1|1')\gamma(2|2')-\gamma(1|2')\gamma(2|1').
\label{cumulant-DEF}
\end{eqnarray}
The index HF means generalized (with non-idempotent occupancies)
Hartree-Fock part of $\gamma_2$. $A$ is the antisymmetrizer.
$\chi$ is the so-called cumulant 2-matrix. (For cumulants, cf. e.g. Ref. 
\cite{Zie9}.) For the diagonal elements, Eq.(\ref{cumulant-DEF}) is read as
\begin{equation}
\rho_2(1,2)=\rho(1)\rho(2)-\gamma(1|2)\gamma(2|1)-u(1,2)
\label{cummPD-DEF}
\end{equation}
with the cumulant PD $u=\text{D}\chi$.
For $\chi$ the contraction sum rule (SR) (\ref{contrSR}) is written as
\begin{equation}
\int d2 \ \chi(1|1',2|2)=\sum_{\kappa}\psi^{}_{\kappa}(1)\nu_{\kappa}
(1-\nu_{\kappa})\psi^{\ast}_{\kappa}(1').
\label{cumcontrSR}
\end{equation}
This contraction SR contains the normalization SR
\begin{equation}
\int d1d2 \ u(1,2)=\text{Tr}\chi=\text{Tr}\gamma(1-\gamma)=\sum_{\kappa}\nu_{\kappa}(1-\nu_{\kappa})
=Nc.
\label{cumnormalizationSR}
\end{equation}
It defines a quantity $c<1$, which is the cumulant PD normalization per 
particle. This quantity is referred to as L\"owdin parameter (because L\"owdin
has asked for the meaning of $\text{Tr}\gamma^2$ \cite{Low}). It vanishes for 
idempotent occupancies and increases with increasing correlation induced 
non-idempotency \cite{Zie1,Zie2,Zie3,Roe}. Note that Eqs.(\ref{cumcontrSR}) and
(\ref{cumnormalizationSR}) are invariant under the exchange 
$\nu_{\kappa}\leftrightarrow (1-\nu_{\kappa})$, what is called
particle-hole symmetry \cite{Rus}.

The spin structure of $\gamma$ is simply
\begin{equation}
\gamma(1|1')=\frac{1}{2}\delta_{\sigma_1,\sigma_1'}\gamma(\mathbf{r}_1|\mathbf{r}_1'),
\quad \gamma(\mathbf{r}_1|\mathbf{r}_1')=\rho f(r), \quad r=|\mathbf{r}_1-\mathbf{r}_1'|,
\quad f(0)=1,
\end{equation}
where the latter expression holds for a homogeneous system with the 
dimensionless 1-matrix $f(r)$. The property $f(0)=1$ makes $\gamma(\mathbf{r}_1|
\mathbf{r}_1')$ correctly normalized. Note that for an spin-unpolarized system 
$\rho(1)=\rho(\mathbf{r}_1)/2$, besides homogeneity makes 
$\rho(\mathbf{r})={\rm const}=\rho$. 

The spin structure of $\gamma_2$ is \cite{Zie10} 
\begin{eqnarray}
\gamma_2(1|1',2|2')&=&\frac{1}{2}
\left(
\delta_{\sigma_1,\sigma_1'}\delta_{\sigma_2,\sigma_2'}-\delta_{\sigma_1,\sigma_2'}\delta_{\sigma_2,\sigma_1'}
\right)
\delta_{\sigma_1,-\sigma_2}\gamma_+(1|1',2|2') \nonumber \\
&+&\frac{1}{2}
\left(
\delta_{\sigma_1,\sigma_1'}\delta_{\sigma_2,\sigma_2'}+\delta_{\sigma_1,\sigma_2'}\delta_{\sigma_2,\sigma_1'}
\right)
\gamma_-(1|1',2|2'),
\label{spingamma2}
\end{eqnarray}
which gives rise to a singlet 2-matrix $\gamma_+$ and a triplet 2-matrix 
$\gamma_-$ with the normalizations $\text{Tr}\gamma_{\pm}=\frac{N}{2}\left( 
\frac{N}{2}\pm 1\right)$. For a homogeneous system the diagonal elements of 
$\gamma_2$ define the spin-dependent dimensionless PD $g$ according to 
$\rho^2g=\rho_2$,
\begin{equation}
g(1,2)=\frac{1}{8}\left( 1-\delta_{\sigma_1,\sigma_2} \right) 
\delta_{\sigma_1,-\sigma_2}
g_+(r_{12})+\frac{1}{8}\left( 1+\delta_{\sigma_1,\sigma_2} \right)g_-(r_{12}), 
\quad r_{12}=|\mathbf{r}_1-\mathbf{r}_2|.
\end{equation}
Here $g_+(r)$ is the singlet PD and $g_-(r)$ is the triplet PD, which are 
normalized as
\begin{equation}
\int d^3r \ \frac{\rho}{2} \left[ g_{\pm}(r)-1 \right]=\pm 1, \quad 
g_{\pm}(\infty)=1.
\label{singtripPDnorm}
\end{equation}
Thus the PDs for spin-parallel ($g_p$) and spin-antiparallel ($g_a$) electron 
pairs can be defined with the help of $g_+$ and $g_-$ as
\begin{equation}
g_p(r)=g_-(r), \quad g_a(r)=\frac{1}{2} \left[ g_+(r)+g_-(r) \right].
\label{parantiparPD}
\end{equation}
They have the following normalizations
\begin{equation}
\int d^3r \ \frac{\rho}{2}\left[ 1-g_p(r) \right]=1, \quad
\int d^3r \ \frac{\rho}{2}\left[ 1-g_a(r) \right]=0, \quad g_{p,a}(\infty)=1.
\label{parantiparPDnorm}
\end{equation}
With these definitions, the spin-summed PD is
\begin{eqnarray}
g(r)=\frac{1}{2}\left[g_a(r)+g_p(r) \right]=
\frac{1}{4}\left[g_+(r)+3g_-(r) \right],  
\int d^3r \ \rho \left[ 1-g(r) \right]=1, \quad g(\infty)=1.
\label{spinsummedPD}
\end{eqnarray}
Note that in Refs.\cite{Zie5,Zie6} the $g_{\pm}(r)$ are defined differently, 
namely with a factor $\frac{1}{2}$, such that we have here $g_{\pm}(\infty)=1$
(whereas in Refs.\cite{Zie5,Zie6} it is $\frac{1}{2}$).

Eqs.(\ref{singtripPDnorm})-(\ref{spinsummedPD}) may be equivalently written in 
terms of the dimensionless cumulant PD $h$ defined by $\rho^2h=u$.
Thus Eq.(\ref{cummPD-DEF}) takes the form 
\begin{equation}
g(1,2)=\frac{1}{4}-\frac{1}{4}\delta_{\sigma_1,\sigma_2}|f(r)|^2-h(1,2),
\label{cummulantPD-DEF}
\end{equation}
from which follows
\begin{equation}
g_p(r)=1-|f(r)|^2-h_p(r), \quad g_a(r)=1-h_a(r).
\label{cumulantparantipar}
\end{equation}
With these definitions the normalization SRs (\ref{parantiparPDnorm}) become
\begin{equation}
\int d^3r \ \frac{\rho}{2} h_p(r)=c, \quad \int d^3r \ \frac{\rho}{2} h_a(r)=0, \quad h_{p,a}(\infty)=0.
\label{cumnormparanti}
\end{equation}
Next singlet/triplet terms $h_\pm$ are defined - analog to 
Eq.(\ref{parantiparPD}) by $h_p=h_-$ and $h_a=\frac{1}{2}[h_++h_-]$. Their 
normalization SRs are
\begin{equation}
\int d^3r \ \frac{\rho}{2}h_{\pm}(r)=\mp c, \quad c=1-\frac{2}{\rho}\int d^3r \ |f(r)|^2,
\quad h_{\pm}(\infty)=0,
\label{cumnorm}
\end{equation}
what follows from Eq.(\ref{cumnormparanti}). 
Analog with Eq.(\ref{spinsummedPD}), the spin-summed cumulant PD is
\begin{equation}
h(r)=\frac{1}{2}[h_a(r)+h_p(r)]=\frac{1}{4}[h_+(r)+3h_-(r)], 
\int d^3r\rho h(r)=c, h(\infty)=0. 
\end{equation}
Similar as for $g_{\pm}$ the definition of $h_{\pm}(r)$ differs from the 
definition used in Refs.\cite{Zie5,Zie6}.

For the $\gamma_{\pm}$ of Eq.(\ref{spingamma2}) the contraction SRs
\begin{equation}
\int d^3r_2 \ \gamma_{\pm}(\mathbf{r}_1|\mathbf{r}_1',\mathbf{r}_2|\mathbf{r}_2)
=\frac{1}{2}\gamma(\mathbf{r}_1|\mathbf{r}_1')\left( \frac{N}{2}\pm 1\right)
\label{gamma2contr}
\end{equation}
hold, what contains the normalization SRs $\text{Tr}\gamma_{\pm}=\frac{N}{2}\left( 
\frac{N}{2}\pm 1\right)$ and agrees together with the spin
structure Eq.(\ref{spingamma2}) also with $\int d2 \gamma_2(1|1',2|2)=\gamma(1|1')(N-1)$
of the contraction SR (\ref{contrSR}). For $\chi_\pm$, the cumulant part of $\gamma_{\pm}$, defined
by $\gamma^{}_{\pm}=\gamma^{\text{HF}}_{\pm}-\chi_{\pm}$, it follows from Eq.(\ref{gamma2contr})
\begin{equation}
\int d^3r_2 \chi_{\pm}(\mathbf{r}_1|\mathbf{r}_1',\mathbf{r}_2|\mathbf{r}_2)=
\mp\frac{1}{2}\left[ \gamma(\mathbf{r}_1|\mathbf{r}_1')-
\frac{1}{2}\int d^3r_2\gamma(\mathbf{r}_1|\mathbf{r}_2)\gamma(\mathbf{r}_2|\mathbf{r}_1')
\right],
\label{chicontract}
\end{equation}
$\text{Tr}\chi_{\pm}=\mp\frac{N}{2}c$. This contraction SR does not contain any
non-size extensive term contrary to the rhs of Eq.(\ref{gamma2contr}). 
The difference $\gamma_\pm^{\rm HF}-\gamma_\pm$ makes the non-size extensively 
normalizable and contractable terms to cancel each other what is the 
prerequisite for the thermodynamic limit. The contraction SR (\ref{chicontract})
allows one for a homogeneous system to calculate the (non-idempotent) momentum 
distribution $n(k)$ by solving a quadratic equation provided that the cumulant 
matrices $\chi_{\pm}$ are known e.g. from perturbation theory [$\chi_{\pm}$ is 
given by linked diagrams, but the RPA-like approximation used in Ref. 
\cite{Zie7} yields only the idempotent $n^0(k)=\Theta(1-k)$] or from the 
natural geminals as solutions of an effective 2-body Schr\"odinger equation.
What the latter means is described in the following. We start with the 
1-matrix $\gamma$, derive from this $\gamma_\pm^{\rm HF}$, the HF part of the 
2-matrix, and use it in the definition $\chi_\pm=\gamma_\pm^{\rm HF}-
\gamma_\pm$.  
%
\section*{Natural orbitals and natural geminals}

For a homogeneous system the natural orbitals of $\gamma(\mathbf{r}_1|
\mathbf{r}_1')$ are plane waves $\varphi_{\mathbf{k}}(\mathbf{r})=
\frac{1}{\sqrt{\Omega}}{\rm e}^{{\rm i}\mathbf{kr}}$ ($\Omega=$ normalization volume) 
and their occupancies give the momentum distribution $n(k)$, resulting thus 
from the Fourier transform of the 1-matrix $f(r)$
\begin{equation}
f(r)=\frac{2}{N}\sum_{\mathbf{k}}n(k){\rm e}^{{\rm i}\mathbf{kr}}, \quad
0<n(k)<1, \quad \sum_{\mathbf{k}}=\int \frac{\Omega d^3k}{(2\pi)^3}.
\end{equation}
$n(k)$ allows one to analyze Compton scattering data. With increasing 
correlation, the quasi-particle weight $z_F=n(1^-)-n(1^+)$ (being 1 for `no 
interaction' or $r_s=0$) decreases. Note that in the more general 
spin-polarized 
case there are two different momentum distributions (for spin-up and spin-down).
But for `no polarization' they coincide. For recent parametrizations of the 
momentum distribution(s) cf. \cite{Zie3,Zie10}. 

With this spectral resolution of the 1-matrix $\gamma$ one can easily 
write down $\gamma_2^{\text{HF}}=\text{A}\gamma\gamma$ and from this follow its 
singlet/triplet components $\gamma_{\pm}^{\text{HF}}$.
Thereby the geminals $\frac{1}{\sqrt{\Omega}} {\rm e}^{{\rm i}\mathbf{KR}}
\frac{1}{\sqrt{\Omega}}{\rm e}^{{\rm i}\mathbf{kr}}$
with $\mathbf{R}=\frac{1}{2}(\mathbf{r}_1+\mathbf{r}_2)$ and $\mathbf{r}=\mathbf{r}_1-\mathbf{r}_2$
appear together with the weight $n(k_1)n(k_2)$ as the probability
of finding two electron momenta $\mathbf{k}_{1,2}=\frac{1}{2}\mathbf{K}\pm\mathbf{k}$, where
$\mathbf{K}=\mathbf{k}_1+\mathbf{k}_{2}$ is the total momentum and
$\mathbf{k}=\frac{1}{2}(\mathbf{k}_1-\mathbf{k}_2)$ is the half relative momentum.
The first factor describes the free-particle center-of-mass motion, whereas
$\frac{1}{\sqrt{\Omega}}{\rm e}^{{\rm i}\mathbf{kr}}=\frac{4\pi}{\sqrt \Omega}
\sum_L {\rm i}^l j_L(k\mathbf{r})Y^{\ast}_L(\mathbf{e}_k)$,
$j_L(k\mathbf{r})=j_l(kr)Y_L(\mathbf{e}_r)$, $L=(l,m_l)$ describes the relative motion for 
$\gamma_2^{\text{HF}}$. Starting with $\gamma_2^{\text{HF}}=\text{A}\gamma
\gamma$ and defining $\gamma_{\pm}^{\text{HF}}$ according to the spin structure 
Eq.(\ref{spingamma2}), it results
\begin{equation}
\gamma^{\text{HF}}_{\pm}=\sum_{\mathbf{k}_{1,2}}n(k_1)n(k_2)
\varphi_{\pm}^0(\mathbf{r}_1,\mathbf{r}_2;\mathbf{k}_1,\mathbf{k}_2) \ 
\varphi_{\pm}^{0\ast}(\mathbf{r}'_1,\mathbf{r}'_2;\mathbf{k}_1,\mathbf{k}_2)
\label{spectralHF}
\end{equation}
with $\varphi_\pm^0=\frac{1}{\sqrt \Omega}{\rm e}^{{\rm i}\mathbf{K}\mathbf{R}}
\frac{4\pi}{\sqrt \Omega}\sum_L^{\pm} {\rm i}^l j_L(k\mathbf{r})
Y^{\ast}_L(\mathbf{e}_k)$ and
\begin{equation}
\text{Tr}\gamma_{\pm}^{\text{HF}}=\sum_{\mathbf{k}_{1,2}}n(k_1)n(k_2)
\left[ 1\pm \delta_{\mathbf{k}_1,\mathbf{k}_2}\right]=\frac{N}{2}\left(\frac{N}{2}\pm1 \right)
\mp\frac{N}{2}c.
\end{equation}
So, the spectral resolution (\ref{spectralHF}) can be written in terms of 
free-electron geminals $j_l(k\mathbf{r})$ and 
an occupancy matrix  
\begin{equation}
\mu_{LL'}(\mathbf{K},k)=\int\frac{d\Omega_k}
{4\pi}Y_L^\ast(\mathbf{e}_k) n(k_1)n(k_2)Y_{L'}(\mathbf{e}_k).
\end{equation}
Its diagonalization yields an $L$-mixing \cite{Zie5}.

With the aim to get finally $\chi_{\pm}=\gamma_{\pm}^{\text{HF}}-\gamma_{\pm}$, 
it is assumed that $\gamma_{\pm}$ has the same form as $\gamma_{\pm}^{\text{HF}}$ in Eq.(\ref{spectralHF})
with the only difference of replacing the free-electron geminals 
$\varphi_{\pm}^0$ or $j_l(kr)$ by interacting-electron geminals 
$\varphi_{\pm}$ or $R_l(r,k)$:
\begin{equation}
\gamma_{\pm}(\mathbf{r}_1\mathbf{r}'_1,\mathbf{r}_2|\mathbf{r}'_2)=
\sum_{\mathbf{k}_{1,2}}n(k_1)n(k_2) 
\varphi_{\pm}(\mathbf{r}_1,\mathbf{r}_2;\mathbf{k}_1,\mathbf{k}_2) \
\varphi_{\pm}^{\ast}(\mathbf{r}'_1,\mathbf{r}'_2;\mathbf{k}_1,\mathbf{k}_2).
\label{spectrapm}
\end{equation}
If these interacting-electron geminals $R_l$ together with the momentum 
distribution $n(k)$ are available, then the PDs $g_{\pm}$ follow from
\begin{equation}
g_{\pm}(r)=2\sum_{L}^{\pm}\frac{2}{N}\sum_{\mathbf{k}}\mu(k)R^2_l(r,k), 
\quad \mu(k)=\frac{2}{N}\sum_{\mathbf{K}}n(k_1)n(k_2)
\label{gwithgeminal}
\end{equation}
with $\frac{2}{N}\sum_{\mathbf{k}} \mu(k)=1$ and $\mu(0)=2^3(1-c)$ .

Assuming that the $R_l$ are scattering solutions of a radial Schr\"odinger 
equation (cf. next Sec.) with a large-$r$ asymptotics (phase shifted compared 
with $j_l$) according to
\begin{equation}
R_l(r,k)\rightarrow \frac{1}{kr}\sin (kr-l\frac{\pi}{2}+\eta_l(k)), \quad 
j_l(kr)\rightarrow \frac{1}{kr}\sin (kr-l\frac{\pi}{2}),
\label{phaseshift}
\end{equation}
the normalization SRs (\ref{singtripPDnorm}) or (\ref{cumnorm}) take the form \cite{Zie5}
\begin{equation}
\frac{2}{\pi}\sum_L^{\pm} \int_0^\infty dk \left [-\frac{\mu(k)}{dk}\right ]\eta_l(k)=\pm c.
\label{normalphase}
\end{equation}
One may compare these SRs with the well-known Friedel SR for point defects in metals with
their screening cloud around an impurity. If the spectral resolutions (\ref{spectralHF})
and (\ref{spectrapm}) are used in $\chi_{\pm}=\gamma_{\pm}^{\text{HF}}-\gamma_{\pm}$, the
contraction SRs (\ref{chicontract}) can be written as \cite{Zie6}
\begin{eqnarray}
\frac{2}{\pi}\sum_L^{\pm} \int_0^\infty dk  
\left [-\frac{\partial 2^3 \mu(2\kappa,k)}{\partial k}\right ]\eta(k)+
b_{\pm}(\kappa)=\pm n(\kappa)[1-n(\kappa)], \nonumber \\  
\mu(K,k)=\int\frac{d\Omega_k}{4\pi}n(k_1)n(k_2).
\label{contractphase}
\end{eqnarray}
The quantity $b_{\pm}(\kappa)$ is defined in the next Section. With 
$\sum_{\boldsymbol{\kappa}}b_{\pm}(\kappa)=0$, the contraction SRs 
(\ref{contractphase}) 
contain the normalization SRs (\ref{normalphase}) as special cases. Thereby 
$\frac{2}{N}\sum_{\boldsymbol{\kappa}}2^3\mu(2\kappa,k)=\mu(k)$ is used. The 
contraction SR (\ref{contractphase}) may be considered as the spectral 
resolution of the normalization SR (\ref{normalphase}). Both 
Eq.(\ref{normalphase}) and Eq.(\ref{contractphase}) are relations between the 
1-body quantity $n(k)$ and the 2-body quantities $R_l(r,k)$ in addition to the 
virial theorem.
%
\section*{The Kimball-Overhauser approach and its generalizations}

Now the question is, where to get from the geminals $R_l(r,k)$. In Refs. \cite{Kim,Over,Gor}
it is intuitevely assumed that they are (at least for the parametrization (\ref{gwithgeminal})
of the PDs) the solution of the radial Schr\"odinger equation
\begin{equation}
\left[-\frac{1}{r} \frac{\partial^2}{\partial r^2}r+\frac{l(l+1)}{r^2}+v_{\pm}(r)-k^2\right]
R_l(r,k)=0,
\label{radSCH}
\end{equation}
which arises from a 2-body Schr\"odinger equation with an effective interaction potential
$v_{\pm}(r)=\frac{1}{r}+v_{\text{scr}}^{\pm}(r)$ possibly different for + (= even $l$)
and - (= odd $l$). This is the Coulomb repulsion between two electrons effectively screened by
the Fermi-Coulomb hole around each electron. This therefore attractive screening potential
comes in the Hartree description \cite{Dav1} from the Poisson equation
$\triangle v_{\text{scr}}(r)=4\pi\rho[1-g(r)]$, which makes the 
approach a self-consistent one: $g^0(r)\rightarrow v_\pm^0(r)
\rightarrow R_l^0(r,k) \rightarrow g^1(r) \rightarrow \cdots$ . The results of 
this approach are in excellent agreement with the quantum-Monte-Carlo
data of Refs. \cite{Cep,Ort}. They have been further improved with the inclusion of exchange
and correlation in Ref. \cite{Dav2}. Meanwhile also the spin-polarized HEG has 
been treated
in this way \cite{Dav3}. These successes say that this approach contains at least some truth,
even if it is not exact and it confirms the above assumption that only the wave functions
change from $\varphi^0$ to $\varphi$ when going from the known $\gamma_{\pm}^{\text{HF}}$ to
the unknown $\gamma_{\pm}$ leaving the occupancy weight unchanged. 

One generalization is the assumption that the PD geminals of Eq.(\ref{radSCH}) 
can be used in the spectral resolution (\ref{spectrapm}) also as 2-matrix 
geminals. This allows one to calculate not only the PDs $g_{\pm}(r)$ but also
the momentum distribution $n(k)$ through the contraction SRs 
(\ref{contractphase}) - at least in principle. The quantity $b_{\pm}(r)$ 
therein is defined by
\begin{eqnarray}
b_{\pm}(\kappa)=-\frac{1}{2}(4\pi)^2\sum_{L,L'}^{\pm}\frac{1}{N^2} 
\sum_{\boldsymbol{\kappa}_{1,2}}
\tilde{v}_{\pm}(\kappa_{12}) \times \nonumber \\ 
\frac{2}{N}\sum_\mathbf{k}
\left[ \mu_{LL'}(2(\boldsymbol{\kappa}-\boldsymbol{\kappa}_1),k)-
\mu_{LL'}(2(\boldsymbol{\kappa}-\boldsymbol{\kappa}_2),k) \right] 
\times \nonumber \\
\left[ \tilde{R}^{}_L(\boldsymbol{\kappa}_1,k)
 \frac{\partial \tilde{R}^{\ast}_{L'}(\boldsymbol{\kappa}_2,k)}{\partial k^2}
-\frac{\partial \tilde{R}^{}_L(\boldsymbol{\kappa}_1,k)}{\partial k^2}
 \tilde{R}^{\ast}_{L'}(\boldsymbol{\kappa}_2,k)
\right],
\end{eqnarray}
where $\tilde{v}_{\pm}(\kappa_{12})$ and ${\tilde R}_L(\boldsymbol{\kappa},k)$ 
are the Fourier transforms of $v_{\pm}(r)$ and $R_L(\mathbf{r},k)$,
respectively. If one starts with a PD $g^0(r)$, then it follows (e.g. in the 
Hartree approximation) the effective interaction potential $v^0_{\pm}(r)$ 
yielding geminals $R^0_l(r,k)$ with their phase shifts $\eta^0_l(k)$ and a PD 
$g^1(r)$. Next with a starting momentum distribution $n^0(k)$ the lhs of 
Eq.(\ref{contractphase}) can be calculated. The result is a quadratic equation.
Its solution gives $n^1(k)$, etc., until self-consistency is reached finally. 
Whether this really works has to be checked. For the "PD to 2-matrix" generalization
in terms of phase shift SRs for normalization and contraction cf. Refs. \cite{Zie5} 
and \cite{Zie6}, respectively. For the generalization of these SRs to the case 
of the spin-polarized HEG cf. Ref. \cite{Zie10}.

Another generalization concerns inhomogeneous systems. It has been already 
discussed in Ref. \cite{Dav1}. Here the problem is alternatively viewed for the
case of an extended system (e.g. a jellium with a crystalline periodic 
background density). Then one has to solve the 2-body Schr\"odinger equation
\begin{equation}
\left \{\sum_{i=1,2}[t(\mathbf{r}_i)+v_{\rm ext}(\mathbf{r}_i)+
v_{\rm H}(\mathbf{r}_i)]+v_\pm(\mathbf{r}_1,\mathbf{r}_2)-
\frac{1}{2}(k_1^2+k_2^2)\right \}\varphi_\pm=0
\label{2SE}
\end{equation}
with the local Hartree potential $v_{\rm H}(\mathbf{r})$ and an appropriately 
screened Coulomb repulsion $v_\pm(\mathbf{r}_1,\mathbf{r}_2)$. The 
self-consistent procedure would have to start 
with a reasonable approximation for the natural orbitals and their occupancies.
>From this follows $\gamma_\pm^{\rm HF}$. From the solution of Eq. (\ref{2SE}) 
[where on the Hartree level, $\rho_2(\mathbf{r}_1,\mathbf{r}_2)$ is needed as 
an input for $v_\pm(\mathbf{r}_1,\mathbf{r}_2)$], follows also $\gamma_\pm$. 
This gives the new PD. But also $\chi_\pm= \gamma_\pm^{\rm HF}-\gamma_\pm$ is 
then available, which - used in Eq. (\ref{cumcontrSR}) - yields new natural 
orbitals and occupancies. 

%
\section*{Summary and outlook}
The Kimball-Overhauser approach for the pair density of the spin-unpolarized 
homogeneous electron gas (HEG) in terms of geminals is revisited from a 
reduced-density-matrix (RDM) point of view. We start with the 
definition of the 2-body RDM and its spectral resolution in terms of natural
geminals. This gives quite naturally the Kimball-Overhauser parametrization of
the pair density in terms of such geminals. Thereby it is assumed that the 
geminals which parametrize the pair density can be used also as natural 
geminals. This assumption has the advantage that also the 1-matrix can be 
calculated as a consequence of the contraction propeties of the 2-matrix. An 
important role plays the size-extensivity of the cumulants, which allows one to
consider the thermodynamic limit. Open questions are:
\begin{itemize}
\item Does the asymptotics Eq.(\ref{phaseshift}) of $R_l(r,k)$ give [via Eq.
(\ref{gwithgeminal})] the correct asymptotics of $g_{\pm}(r)$? If this is not 
the case, can this deficiency be removed by using non-local effective 
interaction potentials ?  
\item Why the lhs of Eqs.(\ref{normalphase}) and (\ref{contractphase}) are note
particle-hole symmetric?
\item Is there a link to the concept of strongly orthogonal geminals? 
\item What are the contraction sum rules for the spin-polarized HEG 
\cite{Zie10}?
\item Can the effective 2-body Schr\"odinger equations (\ref{radSCH}),
(\ref{2SE}) be derived from
the hierarchy of contracted Schr\"odinger equations or from the Bethe-Salpeter equation
or $\cdots$? To what extend is the Kimball-Overhauser approach and its 
generalizations related to a possible pair-density functional theory 
\cite{Zie4} and to 2-body cluster expansions \cite{Pie}?
How to treat finite systems in terms of a screened Coulomb repulsion?
\item What are the peculiarities of $\chi_\pm$ and $v_\pm$ for bondbreaking 
situations, for metals, semiconductors, ferromagnets, antiferromagnets, 
superconductors, off-diagonal-long-range order, ferromagnetic superconductors, 
mixed valence compounds, heavy-fermion systems, non-Fermi liquid behavior, 
quantum criticality, $\cdots$?  
\end{itemize}

In Ref. \cite{Zie11} calculational electronic-structure methods are reviewed
including attempts to generalize density-functional theories (e.g. 
density-matrix functional theory), to further develop the RDM theory 
(contracted Schr\"odinger equations, $N$-representability \cite{Cole}), to make
the accurate quantum-chemical methods (configurational interaction, coupled 
cluster, M\o ller-Plesset) applicable  also to extended 
systems (crystalline solids), and to treat systems with strong electron 
correlations from first principles.  

%
\begin{acknowledgements}
The authors thank P. Gori-Giorgi, J. P. Perdew, M. P. Tosi, R. Asgari, 
V.H. Smith for helpful hints and discussions, P. Fulde and H. Eschrig for 
supporting this work, and J. Cioslowski and K. Pernal for organizing the 
Pomerian Quantum Chemistry and Physics Workshop (Pobierowo, 22 -25 May 2003). 
\end{acknowledgements}

\end{document}